\documentclass[9pt,technote]{IEEEtran}
\def\BibTeX{{\rm B\kern-.05em{\sc i\kern-.025em b}\kern-.08em
    T\kern-.1667em\lower.7ex\hbox{E}\kern-.125emX}}

\usepackage{cite,graphicx,amsmath,amssymb,url,multirow}

\input cyracc.def

\def\cyr{\tencyr\cyracc}

\def\GK{\mbox{\scriptsize GK}}
\def\YS{\mbox{\scriptsize YS}}
\def\s{\mbox{'s}}
\def\boldp{p}
\def\bigp{P}

\def\len{n}
\def\bigl{N}
\def\N{{\mathbb Z}_+}

\def\X{{\mathcal X}}   % alphabet
\def\Z{{\mathbb Z}}

\newcommand{\defn}[0]{\textit}

\begin{document}
\bibliographystyle{IEEEtran}
\title{Prefix Codes for Power Laws with Countable Support}

\author{Michael~B.~Baer,~\IEEEmembership{Member,~IEEE}%
\thanks{M.~B.~Baer is with Electronics for Imaging, 303 Velocity Way, Foster City, CA  94404 USA (e-mail: Michael.Baer@efi.com).}
\thanks{This work has been submitted to the IEEE for possible publication. Copyright may be transferred without notice, after which this version may no longer be accessible.}}
\markboth{IEEE Transactions on Information Theory}{Prefix Codes for Power Laws with Countable Support}

\maketitle

\begin{abstract}
In prefix coding over an infinite alphabet, methods that consider specific distributions generally consider those that decline more quickly than a power law (e.g., Golomb coding).  Particular power-law distributions, however, model many random variables encountered in practice.  For such random variables, compression performance is judged via estimates of expected bits per input symbol.  This correspondence introduces a family of prefix codes with an eye towards near-optimal coding of known distributions.  Compression performance is precisely estimated for well-known probability distributions using these codes and using previously known prefix codes.  One application of these near-optimal codes is an improved representation of rational numbers.
\end{abstract}


\section{Introduction} 
\label{intro}

Consider discrete power-law distributions, those of the form
$$\boldp(i) \sim ci^{-\alpha}$$ for constants $c>0$ and $\alpha>1$,
where $\boldp(i)$ is the probability of symbol~$i$, and $f(i) \sim
g(i)$ implies that the ratio of the two functions goes to $1$ with
increasing~$i$.  Such distributions could be either inherently
discrete or discretized versions of continuous power-law
distributions.

Several researchers in varied fields have, in classic papers ranging
from decades to centuries old, observed power-law behavior for various
discrete phenomena.  These include distribution of
wealth\cite{Pare,Simo}, town and city populations\cite{Auer,Simo},
word frequency\cite{Esto,Zipf,Simo}, numbers of species of a given
genus\cite{Yule,Simo}, and terms in continued
fractions\cite{Gaus,Kuzm}.  More recent papers model various Internet
phenomena\cite{Mitz}.  So active is the topic that several surveys and
popular expositions exist, e.g., \cite{Newm, Mitz, Tale}.

However, there has been relatively little work on lossless compression
of symbols obeying such distributions, in spite of a rich literature
on prefix coding problems\cite{Abr01}.  Exponential-Golomb
codes\cite{Teu} (generalizations of Elias' $\gamma$ code\cite{Eli})
are a good fit for certain power laws\cite{MSW2,MBW}, leading to their
widespread use in compressing video and numerical
data\cite{WSBL,MSW2}.  To the author's knowledge, though, only one
specific infinite-cardinality power-law distribution, the Gauss-Kuzmin
distribution\cite[p.~341]{Knu2}, has been used to judge compression
performance of prefix codes\cite{MaKo,Yoko}.

Here we propose simple codes which not only improve upon existing
codes for encoding symbols distributed according to the Gauss-Kuzmin
distribution --- which applies to coding rational numbers using
continued fractions --- but also efficiently code other common
distributions, such as the zeta distribution with parameter
$2$\cite{Golo2,Kato2}.  We estimate compression performance for dozens
of code/distribution combinations.  For fixed codes, these estimates
are rigorously shown to be precise.

\section{Background, formalization, and motivation}

The most common infinite-alphabet codes are codes that are optimal for
geometric\cite{Golo, GaVV} and geometrically-based\cite{Abr1, ChoG,
MSW, GoMa, BCSV} distributions.  For geometric distributions, these
are known as Golomb codes, and are based on the \defn{unary code} ---
ones terminated by a zero, i.e., a code consisting of codewords the
form $\{1^j0\}$ for $j \geq 0$.  In a Golomb code (G$k$), a unary code
prefix precedes a binary code suffix.  This binary suffix is a
\defn{complete binary code}, in that it has ($k$) codewords of the
same length or length differing by at most one.  For example, the
alphabetic complete binary code of size three that is monotonically
nonincreasing in length is $\{0,10,11\}$, so the Golomb code G$3$ is
$\{00, 010, 011, 100, 1010, \ldots\}$.  If the complete binary code
suffix is of constant length, the overall Golomb code is also called a
Rice code.  Rice codes are used in standards such as
JPEG-LS\cite{WSS}.  Codes that exhibit an efficient coding rate for
power laws, by contrast, are not known to be optimal (excepting those
with finite support and trivial examples for dyadic probability mass
functions).

We restrict ourselves to binary codes and assume that the symbols to
be coded are positive integers.  Thus, an infinite-alphabet source
emits symbols drawn from the alphabet $\X = \{1, 2, 3, \ldots \}$.
(Some applications code the alphabet $\X_0 = \{0, 1, 2, \ldots \}$ or
the alphabet $X_\Z = \{0, -1, 1, -2, 2, \ldots \}$, but any code of
either form can be mapped trivially to a code on $\X$.)  Symbol $i$
has probability $\boldp(i) > 0$, forming probability mass function
$\bigp = \{\boldp(i)\}$.  The source symbols are coded into binary
codewords.  The codeword $c(i) \in \{0,1\}^*$, corresponding to symbol
$i$, has length $\len(i) \in \N$, thus defining length
distribution~$\bigl=\{\len(i)\}$.  An optimal code is one that
minimizes $\sum_{i \in \X} \boldp(i) \len(i)$ with the constraint of a
corresponding code being uniquely decodable, which one is if and only
if the Kraft inequality, $\sum_{i \in \X} 2^{-\len(i)} \leq 1$, is
satisfied.  We can assume without loss of generality that these codes
are prefix codes, that is, codes where there are no two codewords of
the form $c(i)$ and $c(j)=c(i)x$, where $c(i)x$ denotes the
concatenation of strings $c(i)$ and (nontrivial)~$x$.  (In a similar use of
notation, $0^k$ and $1^k$ denote $k$ $0\s$ and $k$ $1\s$,
respectively.  Note also that we use $\lg$ to denote $\log_2$ and
$\ln$ to denote $\log_e$, where $e$ is the base of the natural
logarithm.)

One cannot use the Huffman source coding algorithm\cite{Huff} to find
an optimal code, as one can for a finite source alphabet.  However, it
is sensible that a code over the integers should be \defn{monotonic},
that is, that $\len(i) \geq \len(i+1)$ for all $i \geq 0$.  An
exchange argument easily shows that this is necessary for the code to
be optimal given a distribution for which $\boldp(i) > \boldp(i+1)$
for all~$i$.

Also desirable is for a code to be \defn{alphabetic} or \defn{order
preserving}; that is, if $c(i,j)$ is the $j$th bit of the $i$th
codeword, then $c(i+1,j) < c(i,j)$ only if there is a $k < j$ such
that $c(i+1,k) \neq c(i,k)$.  Alphabetic codes allow the prefix coding
tree to be used as a decision tree, which is useful for search
problems, as in \cite{BeYa,AHK}.  It is also useful for implementation
of arithmetic coding: Because binary arithmetic coding is much faster
than other types of arithmetic coding, a decision tree can reduce an
infinite-alphabet source into a binary source for fast arithmetic
coding, as in \cite{MSW2}.  In addition, order preservation is
necessary for the ordered representation of rational numbers as
integers in continued fractions\cite{MaKo,Yoko}; in this
correspondence we improve upon these representations.

Any valid monotonic prefix code has a (possibly different) alphabetic
prefix code with the same length distribution.  For example, the Elias
$\gamma$ code was first presented in a nonalphabetic version, then
transformed into alphabetic form (as a decision tree) in \cite{BeYa}.
Where there is ambiguity, we will assume use of the alphabetic version
of a code.

Another desirable property is one we call ``smoothness'':

\textit{Definition:} We call $\bigl=\{\len(i)\}$ \defn{$j$-smooth}
if, for every $i>j$, if $\len(i+1)=\len(i+2)$, then $\len(i+1)-\len(i)
\leq 1$, that is, there are no ``jumps'' followed by ``plateaus'';
\defn{weakly smooth} means that it is $j$-smooth for some $j$.  Thus,
for any $j$, a $j$-smooth code includes all weakly smooth codes.
Similarly, $0$-smooth (or \defn{strongly smooth}) codes include all
$j$-smooth (and thus weakly smooth) codes.  Also, we call a
$\bigp=\{\boldp(i)\}$ \defn{$j$-antiunary} if, for every $i>j$,
$\boldp(i) < \boldp(i+1)+\boldp(i+2)$; \defn{antiunary} means that it
is $j$-antiunary for some $j$.

\textit{Observation:} No $j$-antiunary distribution has an optimal code
which is not $j$-smooth.  Thus no antiunary distribution has an
optimal code which is not weakly smooth.

\begin{proof}
Suppose a $j$-antiunary distribution $\bigp$ has an optimal code
with lengths $\bigl$ which is not $j$-smooth.  Then there exists an
$i>j$ such that $\len(i+1)=\len(i+2)$ and $\len(i+1)-\len(i) > 1$.
Consider $\bigl' = \{\len'(i)\}$ for which $\len'(k) = \len(k)$ except
at values $\len'(i)=\len(i)+1$, $\len'(i+1)=\len(i+1)-1$, and
$\len'(i+2)=\len(i+2)-1$.  Clearly $\bigl'$ satisfies the Kraft
inequality and $\sum_i \boldp(i) \len'(i) < \sum_i \boldp(i) \len(i)$,
so $\bigl$ is not optimal.
\end{proof}

Every power law is antiunary, but most previously proposed codes
suitable for power-law distributions are not weakly smooth, so they
could not be optimal solutions.  The proof shows that, when such codes
are applied to antiunary distributions, it is always a simple matter
to improve such a code for use with such a distribution.

For many probability distributions, however, there is no guarantee
that an optimal code would be computationally tractable, let alone
computationally practical for compression applications.  We thus judge
performance of candidate codes by expected bits per coded symbol
rather than by strict optimality.  One of the contributions of this
correspondence is a comparison of various codes for well-known
power-law distributions.

\section{A new family of codes for integers}

\begin{table*}
\centering
\begin{tabular}{r|cl|cl|cl|cl|cl}
&\multicolumn{2}{c}{Code~$-2$} 
&\multicolumn{2}{c}{Code~$-1$} 
&\multicolumn{2}{c}{Code~$0$}
&\multicolumn{2}{c}{Code~$1$} 
&\multicolumn{2}{c}{Code~$2$}
\\
$i$
&$\len_{-2}(i)$&$c_{-2}(i)$ 
&$\len_{-1}(i)$&$c_{-1}(i)$ 
&$\len_{0}(i)$&$c_{0}(i)$ 
&$\len_{1}(i)$&$c_{1}(i)$ 
&$\len_{2}(i)$&$c_{2}(i)$ 
\\
\hline
1&1&0&            1&0 &2&0~0                &3&0~0~0 &4&0~0~00 \\
2&2&10&           3&1~0~0 &3&0~1~0          &3&0~0~1 &4&0~0~01 \\
3&4&11~0~0&       4&1~0~1~0 &3&0~1~1        &4&0~1~0~0 &4&0~0~10 \\
4&5&11~0~1~0&     4&1~0~1~1 &4&10~0~0       &4&0~1~0~1 &4&0~0~11 \\
5&5&11~0~1~1&     5&1~10~0~0 &4&10~0~1      &4&0~1~1~0 &5&0~1~0~00 \\
6&6&11~10~0~0&    5&1~10~0~1 &5&10~1~00     &4&0~1~1~1 &5&0~1~0~01 \\
7&6&11~10~0~1&    6&1~10~1~00 &5&10~1~01    &5&10~0~0~0 &5&0~1~0~10 \\
8&7&11~10~1~00&   6&1~10~1~01 &5&10~1~10    &5&10~0~0~1 &5&0~1~0~11 \\
9&7&11~10~1~01&   6&1~10~1~10 &5&10~1~11    &5&10~0~1~0 &5&0~1~1~00 \\
10&7&11~10~1~10&  6&1~10~1~11 &6&110~0~00  &5&10~0~1~1 &5&0~1~1~01 \\
\end{tabular}
\caption{Five of the Codes introduced here}
\label{codes}
\end{table*}

We propose a family of monotonic, alphabetic, computational efficient,
$0$-smooth codes, starting with the code shown in the center set of
columns ($\len_0(\cdot)$ and $c_0(\cdot)$) of Table~\ref{codes}, which
is defined as
$$c_0(i)=\left\{
\begin{array}{ll}
0b(i-1,3),&i<4\\
1c_0\left(\frac{i-2}{2}\right)0,&i = \{4,6,8,\ldots\}\\
1c_0\left(\frac{i-3}{2}\right)1,&i = \{5,7,9,\ldots\}.
\end{array}
\right.
$$ The term $b(j,k)$ denotes the $(j+1)$th codeword of a complete binary
code with $k$ items, which is order-preserving (alphabetic), with the
first $2^{\lceil \lg k \rceil}-k$ items having length $\lfloor \lg k
\rfloor$ and the last $2k - 2^{\lceil \lg k \rceil}$ items having length
$\lceil \lg k \rceil$.  In
this case, that means that $c_0(1) = 0b(0,3) = 00$, $c_0(2) = 0b(1,3)
= 010$, and $c_0(3) = 0b(2,3) = 011$.  Thus, for example, $c_0(12) =
1c_0(5)0 = 11c_0(1)10 = 110010$.  This is a unary code of length $m$,
followed by a binary digit~$b$ (where $b=0$ or $b=1$), and a binary
code of length $m+b-1$, and is thus straightforward to encode, decode,
and write in the form of an implicit infinite search tree.

This code, like exponential-Golomb codes, is a modification of the
$\gamma$ code.  Whereas the $\gamma$ code has an $m$-bit unary code
followed by a complete binary code for $2^{m-1}$ items, Code~$0$
follows the unary prefix by a complete binary suffix for $3 \cdot
2^{m-1}$ items.  Straightforward extensions of this can be obtained by
modifying the search tree.  We can add a $k$-bit binary
number to each possible codeword --- as in the fourth and fifth set of
columns in Table~\ref{codes} --- extending Code~$0$ in the same manner
that Rice codes extend unary codes, that is,
$$c_k(i) = c_0\left(1+\left\lfloor \frac{i-1}{2^k}
\right\rfloor\right) b((i-1) \bmod 2^k, 2^k)$$ where $k>0$ and
$b\left((i-1) \bmod 2^k, 2^k\right)$ is the $k$-bit representation of
$(i-1) \bmod 2^k$.  Call any of the new extensions Code~$k$.

Another extension, similar to \cite{MSW2} and \cite{Humb1}, involves
first coding with a finite code tree, then, if this initial codeword is all
$1\s$, adding Code~$0$.  If we start as in a unary code and switch to
Code~$0$ after $\kappa$ ones, then let Code~$-\kappa$ denote the
implied code, e.g., Code~$-1$, the second set of columns
($\len_{-1}(\cdot)$ and $c_{-1}(\cdot)$) in Table~\ref{codes}.
Formally, for $k = -\kappa < 0$,
$$c_k(i) = \left\{
\begin{array}{ll}
1^{i-1}0,&i \leq -k\\
1^{(-k)}c_0(i+k),&i > k.
\end{array}
\right.$$

All codes presented here are $0$-smooth (strongly smooth), and can be
coded and decoded using only additions, subtractions, and shifts such
that the total number of operations is proportional to the number of
encoded output bits.

\section{Application}

\begin{table*}
\centering
\begin{tabular}{rll|l|ll|lllll}
&&&\multicolumn{1}{c}{$H$}&\multicolumn{1}{c}{$\bigl^*$}&\multicolumn{1}{c}{Golin}&\multicolumn{1}{c}{Code~$k$}&\multicolumn{1}{c}{{\cyr l}}&\multicolumn{1}{c}{$\gamma$/$\delta$/$\omega$/EG$k$}&\multicolumn{1}{c}{Y}&\multicolumn{1}{c}{G$k$} \\
\hline
\multicolumn{2}{l}{Gauss-Kuzmin}&&$3.43253$&$3.47207$&$3.50705^{(1,2)}$&$\mathbf{3.472346}^{(-1)}$&$3.77915$&$3.50705^{(\gamma)}$&$3.48765$&$\infty^{(\forall k)}$ \\
\hline
\multirow{5}{*}{\rotatebox{90}{Yule-Simon~}}
&$\rho = 1$&&$2.95215$&$2.98136$&$3.^{(1,2)}$&$2.983338^{(-1)}$&$3.17826$&$3.^{(\gamma)}$&$\mathbf{2.98138}$&$\infty^{(\forall k)}$ \\
&$\rho = 1.5$&&$2.17073$&$2.21571$&$\mathit{2.22507}^{(1)}$&$\mathbf{2.230792}^{(-2)}$&$2.32233$&$2.28020^{(\gamma)}$&$2.26031$&$2.85003^{(3)}$ \\ 
&$\rho = 2$&&$1.74685$&$1.83787$&$\mathit{1.84024}^{(1)}$&$\mathbf{1.848484}^{(-4)}$&$1.91747$&$1.94200^{(\gamma)}$&$1.92361$&$2.^{(1)}$ \\
&$\rho = 2.5$&&$1.47629$&$1.62102$&$\mathit{1.62191}^{(1)}$&$\mathbf{1.626668}^{(-5)}$&$1.68947$&$1.74664^{(\gamma)}$&$1.73044$&$2.66666 \ldots^{(1)}$ \\
&$\rho = 3$&&$1.28665$&$1.48534$&$\mathit{1.48563}^{(1,2)}$&$\mathbf{1.488172}^{(-6)}$&$1.54608$&$1.61950^{(\gamma)}$&$1.60550$&$1.5^{(1)}$ \\
\hline
\multirow{2}{*}{\rotatebox{90}{zeta~~}}
&$s = 2$&&$2.36259$&$2.41766$&$2.43310^{(1)}$&$\mathbf{2.417772}^{(-2)}$&$2.53468$&$2.44631^{(\gamma)}$&$2.43042$&$\infty^{(\forall k)}$ \\
&$s = 2.5$&&$1.46525$&1.65431&$\mathit{1.65767}^{(1)}$&$\mathbf{1.658015}^{(-4)}$&$1.70907$&$1.73223^{(\gamma)}$&$1.71963$&$1.94737 \ldots^{(1)}$ \\
&$s = 3$&&$0.97887$&$1.33453$&$\mathit{1.33504}^{(1)}$&$\mathbf{1.336680}^{(-4)}$&$1.36956$&$1.42207^{(\gamma)}$&$1.41389$&$1.36843 \ldots^{(1)}$ \\
\hline
&&&\multicolumn{1}{c}{entropy}&\multicolumn{2}{c}{``designer'' codes \quad \qquad}&\multicolumn{5}{c}{\quad f~~i~~x~~e~~d \qquad c~~o~~d~~e~~s}
\end{tabular}
\caption{Compression (in bits per symbol) and code parameter (where applicable)}
\label{comp}
\end{table*}

Table~\ref{comp} lists various distributions for which no optimal code
is known and estimates, in expected bits per input symbol, of coding
performance using several different codes.  The entropy and the
expected bits per symbol of an optimal code are also estimated.  $H$
denotes the entropy of the distribution ($H(\bigp) = -\sum_i \boldp(i) \lg
\boldp(i)$) and $\bigl^*$ (the expected codeword length of) the
optimal code.  Golin denotes the best Golin code\cite{Gol95}; Code~$k$
denotes the best of the codes introduced here; {\cyr l} denotes the
Levenshtein ({\cyr Levenshte{\u i}n}) code \cite{Leve};
$\gamma$/$\delta$/$\omega$/EG$k$ denotes the best of the Elias
codes\cite{Eli} and exponential-Golomb codes\cite{Teu}, which in these
examples is always the Elias $\gamma$ code (EG$0$); Y denotes
Yokoo's code for the Gauss-Kuzmin distribution\cite{Yoko}; and G$k$
denotes the best Golomb code (with parameter $k$)\cite{Golo}.  These
codes are defined in the cited papers and the definitions are repeated
in the Appendix, which also explains the methods by which the
estimations of bits per symbol are calculated.  In cases for which
there are multiple codes and/or parameters, the best one is chosen and
indicated in superscript.  Note that, as in previous papers on these
and similar codes\cite{Teu,WeVi}, the best code is chosen by its
empirical performance; there appears to be no simple rule for deciding
which code to use.

We show the performance for the overall best fixed code for each
distribution in bold in Table~\ref{comp}, and, if a Golin code is
better, this is in italics.  Note that Golin codes do well for inputs
with rapidly declining probabilities, whereas Yokoo's code and the
codes introduced here have the best results for inverse square
probability mass functions.  However, Golin codes, in being calculated
on the fly, are often impractical, both due to the potential for
rounding errors to lead to coding errors and due to the computational
complexity of the required floating point divisions.

We find that Code~$-1$ is of particular interest as it happens to be
an excellent code
for the Gauss-Kuzmin distribution, defined and well-approximated as follows:
$$\boldp^{\GK}(i) \triangleq - \lg \left[1-\frac{1}{(i+1)^2}\right] \approx
\frac{\lg e}{(i+1)^2}$$ This shows how it is a power law.  The
Gauss-Kuzmin distribution is the one for which to code when
expressing coefficients of continued fractions, as in
\cite{MaKo,KoMa}, in which EG$0$ is proposed for use, and \cite{Yoko},
in which Yokoo's code is proposed.  Code~$-1$ is only about $0.008\%$
worse than the (approximated) optimal code, whereas Yokoo's code is
$0.449\%$ worse and the Elias $\gamma$ code (EG$0$) is $1.007\%$
worse.

Note also that Code~$-2$ is a good code for the
zeta distribution with parameter $s=2$, where the zeta distribution is
defined as
$$\boldp_s^{\zeta}(i) \triangleq \frac{1}{i^s \zeta(s)}$$ and $\zeta$
is the Riemann zeta function $\zeta(s) \triangleq \sum_{i=1}^\infty
i^{-s}$ for $s>1$.  The zeta distribution is used to model several
phenomena including language\cite{Zipf}.  Optimal codes for the zeta
distribution ($s=2$) were considered in Kato's unpublished manuscript
\cite{Kato2}.  In this work, the optimal codeword lengths for the
first ten symbols are shown to lie in ranges of two possible values
for each codeword (or one for the first, which has $\len(1)=1$).  The
codeword lengths of Code~$-2$ all lie within the allowed ranges.
However, we can empirically find better codes, showing that Code~$-2$,
although the best simply described code we know of, is about $0.005\%$
worse than an optimal code.

A third distribution family is that of Yule\cite{Yule} and
Simon\cite{Simo},
$$\boldp_\rho^{\YS}(i) \triangleq \rho B(i,\rho+1) \qquad
\left(\boldp_\rho^{\YS}(i) = \rho
\frac{(i-1)!\rho!}{(\rho+i)!}\right)$$ where $B(i,j)$ is the beta
function, $\rho > 0$, and the right equation applies for integer
$\rho$.  Thus, for example, if $\rho = 1$, then $\boldp(i)=1/i(i+1)$.
Several statistics, from species population to word frequencies,
have been observed to obey a Yule-Simon distribution, most often with
parameter $\rho = 1$\cite{Simo}.  This particular distribution is also
related to continued fractions, being the distribution of the first
coefficient when the number being represented is chosen uniformly over
the unit interval $(0,1)$.  For $\bigp_1^{\YS}$, Yokoo's code is
$0.066\%$ better than Code~$-1$.
 
The estimates in Table~\ref{comp} were calculated based on finite sums
and estimates of the remaining infinite sum.  For fixed codes and for
entropy, these codes are as calculated in the Appendix, and are thus
accurate to the precision given.  The Golin code was estimated based
on the partial code and conditional entropy of the remaining items.
Similarly, optimal expected codeword lengths were estimated using an
optimal code for the partial sum and the entropy of the remaining
items; although not having the same guaranteed accuracy, the results
seem to provide accurate estimates based upon the behavior of coding
truncated probability distributions of increasing size.  In
\cite{LTZ}, it is shown that sequences of such truncated distributions
always have a subsequence converging to the optimal code, providing
theoretical justification for the use of this technique.  Values that
are exactly calculated from infinite sums, rather than estimated, are
indicated by the reduced number of figures (for multiples of $0.1$) or
through ellipses in the case of
$$2.66666 \ldots = \frac{5}{3}, 
1.94737 \ldots = \frac{\zeta(1.5)}{\zeta(2.5)}, \mbox{ and } 
1.36843 \ldots = \frac{\zeta(2)}{\zeta(3)}.$$

These values are exactly known due to being means of Yule-Simon and
zeta distributions, which are known in closed form.  In addition, the
average length of the Elias $\gamma$ code (EG$0$) code for a
Yule-Simon distribution with $\rho=1$ is easily calculated as
\begin{eqnarray*}
\sum_{i=1}^\infty \boldp(i) \len(i) &=& 1+2\sum_{i=1}^\infty \frac{\lfloor \lg i
\rfloor}{i(i+1)} \\
&=&1+2\sum_{j=0}^\infty j \sum_{i=2^j}^{2^{j+1}-1} \left(\frac{1}{i}-\frac{1}{i+1}\right) \\
&=&1+2\sum_{j=0}^\infty j 2^{-j-1} = 3.
\end{eqnarray*}
Golin's algorithms both result in the same code for this distribution,
since the algorithms' conditions result in groupings of probabilities
summing to powers of two.  

Excluding Golin codes, we find that the codes introduced here do
quite well, only failing to improve upon existing fixed codes in one
case, the Yule-Simon distribution with parameter $\rho=1$ ($p(i) =
1/i(i+1)$).  Because Yokoo's code requires computing codewords for
complete binary codes with unequal codeword lengths, however, the
codewords of codes introduced here require less computation to encode
and decode.  For all tested distributions, Yokoo's code and the codes
introduced here are both strict improvements on exponential-Golomb and
Elias codes, confirming that, in practice, strongly smooth codes are
preferable to those lacking this property.

Note that not all known codes for integers were tested here; certain
codes can be ruled out due to the length of the first few codewords
(e.g., Even-Rodeh\cite{EvRo}, Williams-Zobel\cite{WiZo}), whereas
others lack the alphabetic property and/or have significantly higher
computational complexity (e.g., Fibonacci\cite{CaDe0, ApFr}).  In
comparison to feasible codes, the codes introduced here are a notable
improvement.  While not optimal, they can be quite useful in practical
applications.

\appendix

Consider all codes and probability distributions that are monotonic
and for which we can find $\alpha, \beta, \kappa>0, \mu, \xi>0,
\tau>0, \upsilon>0, \phi>0$ such that
$$\len(i) \in \left[ \tau \ln (i+\mu+1) + \alpha, \upsilon \ln (i+\mu) +
\beta \right]$$ and $$\boldp(i) \in
\left[\frac{\phi}{(i+\kappa)^{\xi+1}},
      \frac{\phi}{i^{\xi+1}}\right]$$
for large enough $i \geq i_{\min}$.  Then, for $x>i_{\min}$, we have
\begin{eqnarray*}
\sum_{i=x}^\infty \boldp(i) \len(i) 
&\geq& \int_x^\infty \boldp(i) \len(i-1) di \\
&\geq& \int_x^\infty \frac{\tau\phi \ln(i+\mu) + \alpha\phi}{(i+\kappa)^{\xi+1}} di \\
&\geq& \phi\int_x^\infty \frac{\tau \ln(i+\kappa) + \tau
f_{\min}(x) + \alpha}{(i+\kappa)^{\xi+1}} di \\
&=& \frac{\tau\phi \ln(x+\min(\kappa,\mu)) + \tau\phi\xi^{-1} + 
 \alpha\phi}{\xi(x+\kappa)^{\xi}}
\end{eqnarray*}
where $f_{\min}(x) = \min(\ln (x+\mu)-\ln (x+\kappa),0)$, and
\begin{eqnarray*}
\sum_{i=x}^\infty \boldp(i) \len(i)
&\leq& \int_x^\infty \boldp(i-1) \len(i) di \\
&\leq& \int_x^\infty \frac{\upsilon\phi \ln(i+\mu) + \beta\phi}{(i-1)^\xi} di \\
&\leq& \phi \int_x^\infty \frac{\upsilon \ln(i-1) + \upsilon f_{\max}(x) + \beta}{(i-1)^\xi} di \\
&=& \frac{\upsilon\phi \ln(x+\max(-1,\mu)) + \upsilon\phi\xi^{-1} + 
 \beta\phi}{\xi (x-1)^{\xi}}
\end{eqnarray*}
where $f_{\max}(x) = \max(\ln (x+\mu) + \ln (x-1),0)$, providing upper
and lower bounds to average codeword length using code $\bigl =
\{\len(i)\}$ for probability distribution $\bigp = \{\boldp(i)\}$.
Other distributions (such as Golomb codes) and entropy can be bounded 
similarly.

Such an approach enables us to find estimates with accuracies limited
only by the precision of the partial summations (i.e., round-off
error).  For the probability distributions currently under consideration, we have:

\begin{center}
\begin{tabular}{r|llll}
&$\xi$&$\phi$&$\kappa$ \\
\hline
$\bigp_{\GK}$&$1$&$1$&$\lg e$ \\
$\bigp_\rho^{\YS}$&$\rho$&$\rho$&$\rho \Gamma(\rho+1)$ \\
$\bigp_s^{\zeta}$&$0$&$s-1$&$\zeta^{-1}(s)$ \\
\end{tabular}
\end{center}

In order to find bounds for expected codeword lengths, we should first define
the codes we are using.  Since we only care about codeword lengths, we
use code definitions that apply to $\X$ and have the same lengths
$\bigl$ as the (equivalent but possibly different) original
definitions:
\begin{equation*}
\begin{array}{lrcl}
\mbox{Elias }\gamma&c_{\gamma}(i)&=&\left\{
\begin{array}{ll}
0,&i=1\\
1c_{\gamma}\left(\frac{i}{2}\right)0,&i = \{2,4,6,\ldots\}\\
1c_{\gamma}\left(\frac{i-1}{2}\right)1,&i = \{3,5,7,\ldots\}
\end{array}
\right. \\

\mbox{Elias }\delta&c_{\delta}(i)&=&~c_{\gamma}(1+\left\lfloor \lg i
\right\rfloor) b(i - 2^{\left\lfloor \lg i \right\rfloor},
2^{\left\lfloor \lg i \right\rfloor}) \\

\mbox{Elias }\omega&c_{\omega}(i)&=&\left\{
\begin{array}{ll}
0,&i=1\\
c'_{\omega}(\lfloor \lg i \rfloor)b(i)0,&i>1
\end{array}
\right. \\

\mbox{\cyr l}&c_{\mbox{{\scriptsize {\cyr l}}}}&=&\left\{
\begin{array}{ll}
0,&i=1\\
1c_{\omega}(i-1),&i>1
\end{array}
\right. \\

\mbox{EG$k$}&c_{{\mbox{\scriptsize EG}}k}(i) &=&
~c_{\gamma}\left(1+\left\lfloor \frac{i-1}{2^k} \right\rfloor\right) b((i-1)
\bmod 2^k, 2^k) \\

\mbox{Yokoo}&c_{\mbox{\scriptsize Yok}}(i) &=& \left\{
\begin{array}{ll}
0,&i=1\\
100,&i=2\\
101,&i=3\\
1^{g_i} 00 b(i-2^{g_i}, m_i),&i<q_i\\
1^{g_i} 01 b(i-q_i, 2^{g_i}-m_i),&i \geq q_i
\end{array}
\right.
\end{array}
\end{equation*}
where $c'_{\omega}(i)$ is all but the last bit of $c_{\omega}$,
$g_i \triangleq \lg i$, $m_i \triangleq  (2^{g_i}-(-1)^{g_i})/3$, and $q_i \triangleq
2^{g_i}+m_i$.  Recall that $b(j,k)$ denotes the $(j+1)$th codeword of
a complete binary code with $k$ items.

For these codes, $\alpha, \beta, \mu, \tau>0,
\upsilon>0$ can be
\begin{center}
\begin{tabular}{r|lllll}
&$\alpha$&$\beta$&$\mu$&$\tau$&$\upsilon$ \\
\hline
$\gamma$, Yokoo&$-1$&$1$&$0$&$2 \lg e$&$2 \lg e$ \\
{\cyr l}&$2$&$2$&$-1$&$\lg e$&$2.5 \lg e$ \\[-4pt]
{\scriptsize ($i>1$)}&&&&& \\
Code~$k$&$\alpha_0-k$&$-1-k$&$2+k$&$2 \lg e$&$2 \lg e$ \\[-4pt]
{\scriptsize ($k \leq 0$)}&&&&& \\[-5pt]
{\scriptsize ($i>-k$)}&&&&& 
\end{tabular}
\end{center}
where $\alpha_0 = 1-2\lg 3$.  (Parameters for $\delta$ codes, $\omega$
codes, EG$k$ codes, and Code~$k$ for $k>0$ can be similarly
formulated, but these are unused here, as the $\gamma$ code is clearly
better for all distributions considered.)

For finding the best code within code families with multiple codes ---
such as Code~$k$, EG$k$, and G$k$ (Golomb code $k$, defined in the
main text) --- partial sums can be used to limit the number of codes
tested to a finite number.  For example, these codes have $\len(1)
\rightarrow \infty$ as $k \rightarrow +\infty$, so at some point
$\boldp(1) \len(1)$ will be too large to consider Code~$k$ with
parameters $k > k_{\max}$ for some~$k_{\max}$.  Similarly, as $k
\rightarrow -\infty$, the unary portion of the code can be used for
the partial sum.

Lacking $\alpha, \beta, \mu, \tau, \upsilon$, an obvious lower bound for
$\sum_{i=1}^\infty \boldp(i) \len(i)$ is $\sum_{i=1}^{x-1} \boldp(i)
\len(i)$, but a much more accurate bound can be found via entropy bounding
with a value of $x$ such that $\sum_{i=1}^{x-1} 2^{-\len(i)} =
1-2^{-y_x}$ for some~$y_x$.  For such values, since the code can be
assumed without loss of generality to be monotonic, the codewords can
be assumed to be all the leaves of a subtree rooted at depth~$y_x$.  Since any
normalized tree is subject to the entropy bound $\sum_i \boldp(i)
\len(i) \geq H(\bigp)$, we can normalize to find a useful bound for
the overall code.  Let us first assign 
\begin{eqnarray*}
\sigma_x &\triangleq&\sum_{i=1}^{x-1} \boldp(i), \qquad H_x \triangleq \sum_{i=x}^\infty \boldp(i) \lg \frac{1}{p(i)} \\
H_x^{\mbox{\scriptsize cond}} &\triangleq& \sum_{i=x}^\infty
\frac{\boldp(i)}{1-\sigma_x} \lg \frac{1-\sigma_x}{p(i)} 
= \lg (1-\sigma_x) + \frac{H_x}{1-\sigma_x}
\end{eqnarray*}
where $H_x$ can be lower-bounded by 
as previously described.
Thus, applying the entropy bound to the normalized subtree,
$$\sum_{i=x}^\infty \frac{p(i)}{\sum_{j=x}^\infty \boldp(j)} \left(\len(i) - y_x 
\right) \geq H_x^{\mbox{\scriptsize cond}}$$ so
\begin{eqnarray*}
\sum_{i=1}^\infty \boldp(i) \len(i) &\geq& (y_x + 
  H_x^{\mbox{\scriptsize cond}})(1-\sigma_x) \\
&=&H_x + (y_x + \lg (1-\sigma_x))(1-\sigma_x)
\end{eqnarray*}
This is useful for the codes calculated on the fly, e.g., Golin's codes.

Golin's original approach, \textit{alg1}, starts by finding the minimum
value $k_1$ such that
$$\sum_{i=1}^{2^{k_1}} \boldp(i) > \frac{3-\sqrt{5}}{2} =
0.381966\ldots$$ and assigning the first $2^{k_1}$ inputs code $0
b(i-1,2^{k_1})$.  The algorithm then normalizes the remaining inputs
and finds the minimum value $k_2$ such that
$$\sum_{i=2^{k_1}+1}^{2^{k_1}+2^{k_2}} \boldp_1(i) > \frac{3-\sqrt{5}}{2}
\mbox{ where }
\boldp_1(i) = \frac{\boldp(i)}{1-\sum_{j=1}^{2^{k_1}} \boldp(j)}$$
and assigns the next $2^{k_2}$ inputs code $10
b(i-1-2^{k_1},2^{k_2})$.  Continuing as needed, the algorithm
sequentially finds minimum $k_h$ (given $k_1$ through $k_{h-1}$) such
that
$$S(k_1^h,\bigp) \triangleq \frac{\sum_{i=1+K(h-1)}^{K(h)}
\boldp(i)}{1-\sum_{i=1}^{K(h-1)} \boldp(i)} > \frac{3-\sqrt{5}}{2}$$ where $K(h)
\triangleq \sum_{j=1}^{h} 2^{k_j}$, and assigns code
$$1^{h-1}0b\left(i-1-\sum_{j=1}^{h-1} 2^{k_j}\right)$$ to items
$1+K(h-1)=1+\sum_{j=1}^{h-1} 2^{k_j}$ through $K(h)=\sum_{j=1}^h
2^{k_j}$.  

This top-down approach is quite similar to Shannon-Fano
coding\cite{Shan}, a modification of which results in \textit{alg2},
previously proposed in \cite{BaKo}.  In this case, the the grouping
condition is not the first $k_h$ such that $S(k_1^h,\bigp) >
(3-\sqrt{5})/2$, but the $k_h$ minimizing
$$|S(k_1^h,\bigp)-0.5|$$ that is, the group of a power of two that
results in the most even division between those grouped and those left
ungrouped.  (Note that Shannon-Fano codes use the overall ``best
split'' whereas these codes use the best split that groups items
together in powers of two.)

\ifx \cyr \undefined \let \cyr = \relax \fi

\end{document}